\documentclass[a4paper,conference]{IEEEtran}
\IEEEoverridecommandlockouts

\usepackage{tabularx}
\usepackage{cite}
\usepackage{graphicx}
\usepackage[table, dvipsnames]{xcolor}
\usepackage{color, colortbl}
\usepackage{bbm}
\usepackage{url}
\definecolor{lavenderblue}{rgb}{0.8, 0.8, 1.0}
\definecolor{skyblue}{rgb}{0.53, 0.81, 0.92}
\definecolor{blizzardblue}{rgb}{0.67, 0.9, 0.93}
	\definecolor{aqua}{rgb}{0.0, 1.0, 1.0}
   \graphicspath{{../eps/}}
   \DeclareGraphicsExtensions{.eps}

\usepackage{mathrsfs}
\usepackage{amssymb}
\usepackage{yfonts}
\usepackage{bm}
\usepackage[cmex10]{amsmath}
\usepackage{algorithmicx}
\usepackage{algpseudocode}

\usepackage{algorithm}
\usepackage{graphicx}
\usepackage{array}
\usepackage{color}
\usepackage[caption=false,font=footnotesize]{subfig}
\usepackage{lettrine}
\usepackage{tikz}
\usepackage{verbatim}

{}
\floatname{algorithm}{\textbf{Algorithm}}{}
\ifCLASSINFOpdf

\else

\fi
\begin{document}

\title{Enhancing $6$G Wireless Intelligence: Do LLMs Work for CSI Prediction?\\
\thanks{This work was supported by Institute for Industrial Information Technology (inIT), Technische Hochschule Ostwestfalen-Lippe (TH-OWL), $32657$ Lemgo, Germany.}
}

\author{\IEEEauthorblockN{Mohsen Kazemian}
\IEEEauthorblockA{\textit{Institute Industrial IT (inIT)} \\
\textit{Technische
Hochschule OWL (TH-OWL)}\\
Lemgo, Germany \\
mohsen.kazemian@th-owl.de}

\and
\IEEEauthorblockN{Jürgen Jasperneite}
\IEEEauthorblockA{\textit{Institute Industrial IT (inIT)} \\
\textit{Technische
Hochschule OWL (TH-OWL)}\\
Lemgo, Germany\\
juergen.jasperneite@th-owl.de}
}


\maketitle

\begin{abstract}
In high-mobility $6$G scenarios, rapidly time-varying channels lead to very short coherence times, which makes conventional pilot-based channel state information (CSI) estimation approaches prone to outdated information or excessive pilot overhead. Therefore, channel prediction becomes essential in such dynamic wireless systems. To address this challenge, large language models (LLMs) are emerging learning frameworks that have recently attracted attention for CSI prediction due to their strong sequence modeling capability and ability to generalize across different environments. This paper proposes an LLM-based framework for channel prediction in high-mobility orthogonal time frequency space (OTFS) communication systems. In this work, we develop a physics-aware LLM-based predictor that learns the temporal evolution of OTFS channel coefficients from historical channel observations while incorporating mobility-related physical descriptors (e.g., maximum Doppler frequency) to achieve accurate prediction of future channel states in rapidly time-varying environments. The effectiveness of the proposed framework is evaluated through extensive simulations under user velocities ranging from $100$ to $500$ km/h. Numerical results show that the proposed method consistently achieves lower normalized mean square error (NMSE) compared with both classical deep learning predictors and LLM-based predictors without physical channel descriptors. These results demonstrate the advantage of integrating mobility-related channel knowledge with LLM-based sequence modeling for channel prediction in highly dynamic OTFS systems.
\end{abstract}


\IEEEpeerreviewmaketitle

\section{Introduction}

Sixth-generation ($6$G) wireless networks are expected to support a wide range of high-mobility applications, including satellite communications, unmanned aerial vehicles (UAVs), aircraft, and high-speed trains. In such scenarios, conventional orthogonal frequency-division multiplexing (OFDM) systems suffer from severe performance degradation due to inter-carrier interference (ICI) caused by large Doppler shifts, as well as limited spectral efficiency \cite{jamm}, \cite{gray}. To address these challenges, orthogonal time frequency space (OTFS) modulation has recently emerged as a promising waveform for high-mobility communications. Unlike OFDM, OTFS maps information symbols onto the delay–Doppler (DD) domain, allowing the system to exploit the full time–frequency (TF) diversity of the wireless channel even in uncoded systems.

However, the rapidly varying nature of wireless channels in high-mobility environments leads to very short channel coherence times. As a result, conventional pilot-based channel estimation techniques require frequent pilot transmissions to maintain accurate channel state information (CSI) at the receiver, which significantly reduces spectral efficiency. On the other hand, reducing the number of pilots leads to outdated CSI due to channel aging. Therefore, to obtain accurate and real-time CSI, studying prediction techniques in OTFS systems is essential for reliable communication in high-mobility $6$G applications.

Since conventional channel estimation techniques are not well suited for rapidly time-varying channels, this paper focuses on channel prediction methods. These approaches can be broadly categorized into two main groups: model-based methods and AI-based methods. The first group relies on physical parameters of the wireless propagation environment to characterize channel dynamics. To the best of our knowledge, \cite{FAROTFS} is the first study that works in this area and formulates OTFS channel prediction using a functional autoregressive (FAR) model. In this approach, the nonlinear functional relationship between successive OTFS frames is learned while jointly estimating the innovation variance of the channel process. Several earlier works assume a linear temporal relationship between consecutive channel states \cite{arman}, which cannot accurately capture the nonlinear evolution of wireless channels under high-mobility conditions. Although model-based methods provide strong physical interpretability and low computational complexity, their performance strongly depends on the accuracy of the assumed channel model and the estimation of physical parameters such as Doppler spread, delay spread, and noise statistics. In practice, these parameters may vary significantly across environments and mobility conditions, making model-based predictors sensitive to parameter mismatch. Moreover, simplified assumptions used in analytical models may fail to capture the complex nonlinear dynamics of wireless channels in highly time-varying scenarios.

On the other hand, AI-based channel prediction methods rely on learning temporal channel patterns from datasets rather than explicitly modeling physical channel parameters. These methods can be further divided into two subgroups: approaches based on conventional deep learning models \cite{2aidriven} and those based on large language models (LLMs). Among the deep learning-based approaches, a number of works exploit the structured representation of OTFS channels in the DD domain. For example, \cite{aione} proposes a hybrid channel estimation and prediction scheme for massive MIMO–OTFS systems in which uplink channels are first estimated using a basis expansion model (BEM) and then represented using Slepian sequences whose coefficients are extrapolated using orthogonal polynomial fitting to predict future channels. Although this approach improves prediction accuracy and reduces pilot overhead, it relies heavily on predefined basis models and may struggle to capture complex nonlinear channel dynamics.
Another recent line of work employs deep neural networks for this target. For instance, a hybrid convolutional neural network and transformer architecture has been proposed in \cite{aitwo}, in which a CNN extracts compact spatial features from the DD channel representation, while a transformer captures temporal dependencies across frames using causal attention. Experimental results under extreme mobility conditions (e.g., $500$ km/h) show improved prediction accuracy compared to conventional baselines. 

In contrast to conventional channel prediction approaches that rely on historical CSI observations, the CVAE$4$CP framework in \cite{cvae} formulates OTFS channel prediction as a conditional generative modeling problem driven by channel generation parameters. Specifically, a conditional variational autoencoder (CVAE) learns the distribution of DD channel coefficients conditioned on physical system and mobility parameters, enabling the generation of future CSI realizations while capturing channel uncertainty through a low-dimensional latent representation.

Nevertheless, such AI-based models typically require large training datasets and are often trained separately for each environment, limiting their ability to generalize across heterogeneous wireless scenarios. Motivated by the recent success of LLMs in sequence modeling and cross-domain generalization, LLM-based approaches are emerging as a promising direction for wireless channel prediction. Besides, in the time-series forecasting literature, transformer-based models such as PETformer \cite{pet} have been proposed to capture temporal dependencies between sequential observations, and such architectures have demonstrated strong potential for modeling sequential data across multiple domains, including wireless communication systems.

Recent advances in sequence modeling have motivated the use of transformer-based architectures for wireless channel prediction. In this direction, the linear-based lightweight transformer (LinFormer) framework in \cite{linformer} employs an encoder-only transformer architecture for CSI prediction and replaces the self-attention mechanism with a lightweight time-aware multilayer perceptron (TMLP). In conventional transformers, self-attention captures long-range temporal dependencies by computing similarity-based interactions between all tokens in the input sequence. Specifically, given the query ($Q$), key ($K$), and value ($V$) matrices derived from the input sequence $X$, self-attention computes
\begin{equation}
\text{Attention}(Q, K, V) = \text{Softmax}\left(\frac{QK^{T}}{\sqrt{d}}\right)V,
\end{equation}
where $d$ is the feature dimension and the softmax operation produces attention weights representing the relevance between tokens. LinFormer replaces this attention mechanism with a TMLP block that mixes temporal information through learned linear transformations as $O = \tilde{W}\,\mathrm{ReLU}(WX)$,
where $W$ and $\tilde{W}$ are learnable weight matrices, and $O$ is the output representation after temporal mixing. This design significantly reduces computational complexity while preserving the ability to capture temporal dependencies. However, the TMLP uses fixed mixing weights that are independent of the channel dynamics and does not explicitly incorporate physical channel parameters such as Doppler or delay spread, which may limit its ability to adapt to rapidly changing wireless environments.

Motivated by the strong sequence modeling capability of LLMs, several recent works have explored adapting pretrained language models for wireless channel prediction tasks. Since the application of LLMs to channel prediction is still relatively new, in the following, we provide a comprehensive review of the most recent methods in this emerging research direction.
\begin{table*}[t]
\centering
\caption{Comparison of Channel Prediction and a Time-Series Forecasting Method Discussed in the Literature Review}
\label{tab:comparison_all_ieee}
\renewcommand{\arraystretch}{1.15}
\setlength{\tabcolsep}{2.5pt}
\footnotesize
\begin{tabular}{|c|l|l|l|l|l|l|l|l|}
\hline
\textbf{\#} 
& \textbf{Method} 
& \textbf{System} 
& \textbf{Input} 
& \textbf{Model} 
& \textbf{Arch.} 
& \textbf{Attn.} 
& \textbf{Fusion / Conditioning} 
& \textbf{Key Feature} \\
\hline
1 & FTAlign-LLM 
& mMIMO OFDM 
& CSI (F+D) 
& GPT-2 
& Dec 
& Self 
& Freq--delay align 
& Struct.-aware LLM \\
\hline
2 & CSI-ALM / Light 
& mMIMO 
& CSI patches 
& GPT-2 
& Dec 
& Self 
& Semantic prompts 
& Few-shot general. \\
\hline
3 & LinFormer 
& mMIMO 
& Past CSI 
& Linear 
& Enc 
& None 
& -- 
& Ultra-fast infer. \\
\hline
4 & ISAC-LLM 
& OFDM ISAC 
& Comm + sens. CSI 
& GPT-2 
& Dec 
& Self 
& ConvLSTM + cross 
& Exploits scatterers \\
\hline
5 & FAS-LLM 
& OTFS LEO 
& Comp. DD tensor 
& LLaMA 
& Dec 
& Self 
& PCA + phase rec. 
& Scales DD grids \\
\hline
6 & PETformer 
& LTSF 
& Patched TS 
& Transformer 
& Enc 
& Self 
& Placeholders + MSI 
& Future coupling \\
\hline
7 & LLM4CP 
& mMIMO 
& CSI tensors 
& GPT-2 
& Dec 
& Self 
& CSI attention 
& LLM baseline \\
\hline
8 & FAR-OTFS 
& OTFS 
& BEM + covar. 
& FAR 
& AR 
& None 
& Doppler + TARP 
& Interpretable model \\
\hline
9 & CVAE4CP 
& OTFS 
& DD + phys. par. 
& CVAE + flow 
& Enc--Dec 
& None 
& Conditional latent 
& Uncertainty-aware \\
\hline
\end{tabular}
\end{table*}

One representative approach is LLM$4$CP \cite{llmcp}, which employs a pretrained GPT-$2$ architecture for channel prediction in MISO-OFDM systems. In this framework, historical uplink CSI is first transformed into both frequency-domain and delay-domain representations, and the real and imaginary parts are concatenated to form real-valued feature tokens. These tokens are then mapped to the embedding space of GPT-$2$ and processed by a pretrained transformer decoder to predict future CSI. Formally, the prediction process can be written as
\begin{equation}
\hat{H}_{t+1:t+N_F} = f_{\theta}\left(H_{t-N_P+1:t}\right),
\end{equation}
where $H_{t-N_P+1:t}$ denotes the past CSI sequence of length $N_P$, $N_F$ is the prediction horizon, and $f_{\theta}$ represents the LLM-based predictor with parameters $\theta$. However, \cite{llmcp} works on OFDM, which is not appropriate for high-mobility scenarios.

Another work in \cite{isac} introduces sensing-assisted LLM-based channel prediction, where additional sensing information is integrated with communication CSI to improve prediction accuracy. In this method, called integrated sensing and communication (ISAC)-LLM, historical communication CSI and sensing CSI are processed through convolutional long short-term memory (ConvLSTM) networks to extract spatiotemporal features. These features are then fused using cross-attention mechanisms before being fed into a GPT-$2$ backbone for sequence prediction. Although the use of sensing information improves robustness and prediction accuracy, the approach assumes the availability of additional sensing data and introduces additional system complexity.

A more recent framework, frequency-temporal alignment with
LLM (FTAlign-LLM), proposes a frequency–temporal alignment architecture for CSI prediction in cell-free massive MIMO systems \cite{ftalign}. In this approach, CSI is first represented in both frequency and delay domains, and multi-scale CSI attention (MSCA) networks are employed to extract spatial and temporal channel features. The resulting features are then fused using a frequency–temporal feature fusion (FTFF) module, which performs cross-attention between delay-domain and frequency-domain representations before feeding the aligned features into a pretrained GPT-$2$ model. This architecture enables the model to capture both local multipath structure and global spectral correlations. Nevertheless, the framework requires complex preprocessing modules and relies on multiple neural components, which increases the overall model complexity and training cost.

The authors in \cite{fasllm} introduce fluid antenna systems
(FAS)-LLM, which combines FAS with LLM-based channel prediction in OTFS-enabled satellite communications. In this framework, the high-dimensional OTFS DD channel tensor is first compressed using a two-stage procedure consisting of reference port selection and separable principal component analysis (PCA). The compressed representation is then used as input to a pretrained LLaMA model with parameter-efficient fine-tuning to forecast future channel states. After prediction, the full DD channel is reconstructed using the PCA basis and deterministic phase relationships between antenna ports. While this approach significantly reduces input dimensionality and enables efficient forecasting, the compression stage may discard useful channel information and requires additional reconstruction steps. 

To bridge the modality gap between numerical CSI data and the linguistic representations learned by pretrained large language models, CSI-ALM \cite{bridge} introduces a cross-modal fusion module that aligns CSI embeddings with the language feature space by maximizing the cosine similarity between CSI representations and word embeddings. However, this approach relies on additional semantic alignment and knowledge-distillation mechanisms, which increase architectural complexity and may limit its scalability and applicability in practical wireless channel prediction systems.

Table $1$ shows a comparison of the channel prediction techniques discussed above, highlighting their key characteristics and including time-series forecasting method in \cite{pet}.

The aforementioned limitations of existing channel prediction approaches motivate the investigation of new methods that can effectively capture the dynamics of OTFS channels while remaining robust across heterogeneous environments. In this context, recent advances in LLMs have demonstrated strong capabilities in modeling complex sequential dependencies and generalizing across domains. Although LLMs have recently attracted attention in wireless communications, their application to OTFS channel prediction remains largely unexplored and is still in its early stages. Motivated by these observations, in this work we propose an LLM-based channel prediction framework for OTFS systems that operates directly in the DD domain. Therefore, we develop a physics-aware LLM-based framework that learns the temporal evolution of the channel by jointly utilizing historical OTFS channel coefficients and mobility-related physical descriptors, such as Doppler-related information. By combining data-driven sequence modeling with physically meaningful channel descriptors, the proposed approach enables more accurate prediction of future channel states under rapidly time-varying conditions. The main contributions of this work are summarized as follows:

$\bullet$ We propose a physics-aware LLM-based framework for OTFS channel prediction that integrates historical channel observations with physical channel descriptors. By incorporating appropriate mobility-related parameters into the learning process, the proposed approach enables the LLM to better capture the temporal evolution of rapidly time-varying OTFS channels.

$\bullet$ We investigate the impact of different physical parameters
as auxiliary inputs for channel prediction in high-mobility environments. In particular, we show that mobility-related descriptors such as Doppler frequency provide more
informative guidance for channel dynamics, while the signal-to-noise ratio (SNR) is less effective, because it does not directly capture the
temporal evolution of the wireless channel. 



$\bullet$ We conduct extensive simulations under high-mobility conditions to evaluate the proposed framework. The results demonstrate that incorporating physically meaningful channel descriptors significantly improves prediction accuracy compared with conventional deep learning predictors and LLM-based predictors that rely solely on historical channel observations.

\section{System Model}
In this paper, we model the wireless channel as a continuous Doppler spread channel (CDSC) with channel length $L$ and Doppler spectrum following Jakes’ model. To characterize the temporal variation of the CDSC channel while reducing the number of channel parameters, we employ the complex exponential (CE)-BEM, which represents the time-varying channel using a reduced set of basis coefficients while effectively capturing the Doppler-induced dynamics. The channel impulse response (CIR) coefficient can be expressed as
\begin{equation}
{h}[t,l] = \sum_{q=0}^{Q-1} {c}_q[l] \phi_q(t) + e_{\text{BEM}}[t,l],
\end{equation}
where ${h}[t,l]$ denotes the fading coefficient of the $l$-th propagation path at the 
$t$-th OTFS frame (i.e., observation window) with $l \in\{0,...,L-1\}$, $Q$ is the order of the CE-BEM model, $c_q[l]$ is the coefficient associated with the $l$-th delay tap, $\mathbf{c}_q \in \mathbb{C}^{L \times 1}$ is the CE-BEM coefficient vector corresponding to the $q$-th basis function  
$\phi_q(t)$, and $e_{\text{BEM}}[t,l]$ denotes the BEM modeling error corresponding to the $l$-th delay tap in the $t$-th OTFS frame. The received signal corresponding to the $t$-th OTFS frame can be expressed as $y_t = \mathbf{h}_t^{\top}\mathbf{x}_t + n_t$ ,
where $\mathbf{h}_t \in \mathbb{C}^{L}$ denotes the channel coefficient vector at $t$-th frame, $\mathbf{x}_t$ represents the transmitted signal vector, and $n_t$ denotes additive white Gaussian noise (AWGN).
\section{Proposed Method}
In this section, we propose a framework that predicts future channel coefficients based on historical channel observations and physical dynamic information describing the propagation environment. The overall architecture consists of four main components: a preprocessing module, a feature embedding module, an LLM backbone, and an output module.

Given the channel observations from the previous $N_P$ OTFS frames, the historical channel sequence can be written as
$\bm{\mathcal{H}}_t \triangleq \{\mathbf{h}_{t-N_P+1}, \ldots, \mathbf{h}_t\}$, where $\mathbf{h}_t \triangleq \big[h[t,0],\, h[t,1],\, \ldots,\, h[t,L-1]\big]^{\top} \in \mathbb{C}^{L \times 1}$. In addition to the channel coefficients, we also incorporate a physical dynamic descriptor that reflects the temporal variation of the wireless environment. This descriptor may include parameters such as Doppler spread, user velocity, or SNR. Let the physical dynamic sequence be denoted by $\bm{\mathcal{D}}_t \triangleq \{ {\mathbf{d}_{t-N_P+1}, \ldots, \mathbf{d}_t}\}$,
where $\mathbf{d}_t$ represents the physical dynamic features at the $t$-th OTFS frame. The objective of channel prediction is to estimate the channel coefficients of the next $N_F$ OTFS frames based on historical channel observations and the physical dynamic information, i.e., $\{\hat{\mathbf{h}}_{t+1}, \ldots, \hat{\mathbf{h}}_{t+N_F}\}
= \bm{\mathcal{F}}_{\Theta}(\bm{\mathcal{H}}_t, \bm{\mathcal{D}}_t)$, where $\bm{\mathcal{F}}$ denotes the proposed LLM-based predictor with trainable parameters ${\Theta}$.
 
\subsection{Preprocessing Module}\label{bb34}

The preprocessing module prepares the historical channel coefficients and physical dynamic features for input into the neural network. Since neural networks operate on real-valued data, the complex channel coefficients are converted into real representations by separating their real and imaginary parts.
Specifically, each channel vector $\mathbf{h}_t$ is transformed as $\mathbf{x}_t \triangleq \mathrm{concat}\big(\mathrm{Re}(\mathbf{h}_t),\, \mathrm{Im}(\mathbf{h}_t)\big) \in \mathbb{R}^{2L}$. To improve training stability, the input data are normalized using batch statistics. The historical channel sequence is then arranged into a matrix $\mathbf{X}_H \in \mathbb{R}^{2L \times N_P},$where each column corresponds to one OTFS frame.
Similarly, the physical dynamic sequence is represented as
$\mathbf{X}_D \in \mathbb{R}^{K_d \times N_P},$
where $K_d$ denotes the dimension of the dynamic descriptor.

\subsection{Feature Embedding}
To extract useful temporal features from the input sequences, we first project the channel and dynamic inputs into a common embedding space. The channel feature matrix $\mathbf{X}_H$ is mapped through a fully connected layer to obtain $\mathbf{E}_H=\mathrm{F}C_H(\mathbf{X}_H)$. Similarly, the dynamic descriptor is embedded as $\mathbf{E}_D = \mathrm{F}C_D(\mathbf{X}_D)$. The two feature representations are then combined through feature fusion to obtain $\mathbf{E} = \mathbf{E}_H + \mathbf{E}_D$. To preserve the temporal ordering of OTFS frames, positional encoding is added to the fused embedding.

\subsection{LLM Backbone}
After embedding, the input sequence is processed by a pre-trained LLM. We adopt a transformer-based LLM backbone such as GPT-$2$ due to its strong ability to capture long-range dependencies in sequential data. The embedded sequence is treated as a series of tokens and fed into the LLM layers as
$\mathbf{Z} = \mathrm{LLM}(\mathbf{E})$. During training, most parameters of the pre-trained LLM are frozen to retain the general sequence modeling capability, while only a small subset of parameters, such as layer normalization and embedding layers, are fine-tuned to adapt the model to the channel prediction task.

\subsection{Output Module}
The output module converts the LLM features into predicted future channel coefficients. First, a fully connected layer maps the LLM output to the target dimension as
$\hat{\mathbf{X}} = \mathrm{FC}_{\text{out}}(\mathbf{Z})$.

The predicted real-valued representation is then rearranged to reconstruct the complex channel coefficients as
\begin{equation}
\hat{\mathbf{h}}_{t+i} =
\hat{\mathbf{x}}_{t+i}^{(1:L)}
+ j\,\hat{\mathbf{x}}_{t+i}^{(L+1:2L)},
\quad i = 1, \ldots, N_F,
\end{equation}
where $\hat{\mathbf{{x}}}_{t+i} \in \mathbb{R}^{2L}$ denotes the predicted real-valued channel representation at frame $t+i$. The main steps of the proposed framework are summarized in Algorithm $1$.
\begin{algorithm}[t]
\caption{LLM-based OTFS Channel Coefficient Prediction}
\label{alg:llm_otfs_prediction}
\begin{algorithmic}[1]
\Require Historical channel coefficients $\bm{\mathcal{H}}_t=\{\mathbf{h}_{t-N_P+1},\dots,\mathbf{h}_t\}$; historical physical dynamic inputs $\mathcal{D}_t=\{\mathbf{d}_{t-N_P+1},\dots,\mathbf{d}_t\}$
\Ensure Predicted future channel coefficients $\hat{\bm{\mathcal{H}}}_{t+1:t+N_F}=\{\hat{\mathbf{h}}_{t+1},\dots,\hat{\mathbf{h}}_{t+N_F}\}$

\State Convert complex channel vectors $\mathbf{h}_i$ into real-valued representations
\State Form channel feature matrix $\mathbf{X}_H=[\mathbf{x}_{t-N_P+1},\dots,\mathbf{x}_t]$
\State Form physical feature matrix $\mathbf{X}_D=[\mathbf{d}_{t-N_P+1},\dots,\mathbf{d}_t]$

\State Embed channel features: $\mathbf{E}_H=\text{FC}_H(\mathbf{X}_H)$
\State Embed physical dynamic features: $\mathbf{E}_D=\text{FC}_D(\mathbf{X}_D)$

\State Fuse embeddings: $\mathbf{E}=\mathbf{E}_H + \mathbf{E}_D$
\State Add positional encoding: $\mathbf{E}_{in}=\mathbf{E}+\mathbf{E}_{pos}$

\State Feed $\mathbf{E}_{in}$ into the LLM backbone:
\[
\mathbf{Z}=\text{LLM}(\mathbf{E}_{in})
\]

\State Map LLM output to future channel representation:
\[
\hat{\mathbf{X}}=\text{FC}_{out}(\mathbf{Z})
\]

\For{$f=1$ to $N_F$}
\State Reconstruct complex channel coefficients:
\[
\hat{\mathbf{h}}_{t+f}=\hat{\mathbf{x}}_{t+f}^{(1:L)}+j\,\hat{\mathbf{x}}_{t+f}^{(L+1:2L)}
\]
\EndFor

\State \textbf{return} $\hat{\bm{\mathcal{H}}}_{t+1:t+ {N_F}}$
\end{algorithmic}
\end{algorithm}
\begin{table}
\centering
\caption{Simulation Parameters}
\label{tab:simulation_parameters}
\begin{tabular}{ll}
\hline
\textbf{Parameter} & \textbf{Value} \\ 
\hline
Carrier frequency $f_c$ & $3$ GHz \\
Subcarrier spacing $\Delta f$ & $15$ kHz \\
System bandwidth & $8.64$ MHz \\
Number of resource blocks & $48$ \\
Transmit power $P_T$ & $MN$ \\
Modulation scheme & $16$-QAM \\
Detection method & Log-likelihood ratio (LLR) \\
Number of clusters & $21$ \\
Rays per cluster & $20$ \\
Channel length $L$ & $10$ \\
Number of non-zero taps $K$ & $4$ \\
User velocity & $100$--$500$ km/h \\
\hline
\end{tabular}
\end{table}

\section{Simulation Results}

In this section, we evaluate the performance of the proposed method using synthetic time-varying channels generated by the QuaDRiGa simulator following the $3$GPP urban macro (UMa) model under NLOS conditions. The base station employs a dual-polarized uniform planar array with $N_h = N_v = 4$, while the user terminal uses a single omnidirectional antenna. We consider an OTFS system with $M=64$ delay and $N=16$ Doppler bins. Table II
summarizes the remaining simulation parameters used in the experiments. The generated synthetic dataset includes $10,000$ samples, where $N_{tr}=8,000$, $N_{va}=1,000$ and $N_{te}=1,000$ samples are used for training, validation, and testing, respectively. Finally, $N_P=16$, $N_F=4$, and the considered velocities lie in the range of $100–500$ km/h.

During the training phase, the proposed predictor is trained using the generated channel dataset to learn the temporal evolution of OTFS channel coefficients. Each training sample consists of two parts: ($1$) the historical channel coefficients from the previous $N_P$ OTFS frames and ($2$) the corresponding physical dynamic descriptor of the wireless environment. In this work, the physical descriptor is represented by the maximum Doppler frequency, which reflects the mobility-induced temporal variation of the channel. The target output for each sample is the channel coefficients of the subsequent $N_F$ OTFS frames. Using these input–output pairs, the LLM learns a mapping between past channel observations and future channel states while incorporating the physical dynamics of the propagation environment. 

Let $\mathrm{NMSE} \triangleq \mathbb{E}\left(
\frac{\|\mathbf{H}_{DD} - \hat{\mathbf{H}}_{DD}\|_F^2}
{\|\mathbf{H}_{DD}\|_F^2}
\right)$, where $\mathbf{H}_{DD}$ represents the ground-truth DD channel matrix and $\hat{\mathbf{H}}_{DD}$ denotes the predicted channel matrix obtained from the model. The operator $\|\cdot\|_F$ denotes the Frobenius norm and $\mathbb{E}(\cdot)$ represents the expectation over all test samples.
The model parameters are optimized by minimizing the NMSE between the predicted channel coefficients and the ground-truth channel coefficients over the training dataset. After the training process converges, the learned model is evaluated using an independent test dataset that is not seen during training. For each test sample, the historical channel coefficients and their corresponding Doppler descriptors from the previous $N_P$ OTFS frames are fed into the trained LLM predictor. The model then generates estimates of the channel coefficients for the next $N_F$ frames. These predicted channel realizations are compared with the true channel coefficients obtained from the simulator. Finally, the NMSE between the predicted and ground-truth channels is computed over all test samples to quantify the channel prediction accuracy.

Fig. $1$ illustrates the NMSE performance versus user velocity for various methods. As the velocity increases, the prediction accuracy of all methods degrades due to the reduced temporal correlation of the channel in high-mobility environments. Nevertheless, the proposed physics-aware framework consistently achieves the lowest NMSE across the entire velocity range. For example, at a velocity of $450$~km/h, the proposed Doppler-aware predictor attains an NMSE of approximately $4\times10^{-3}$, while the variant using SNR reaches about $1.1\times10^{-2}$ and the ablation model yields roughly $4\times10^{-2}$. Here, the ablation model corresponds to an LLM-based predictor that relies only on historical channel observations without incorporating physical channel descriptors, which is conceptually similar to existing LLM-based CSI prediction approaches such as \cite{llmcp}, but adapted to the OTFS setting. These results show that although incorporating SNR provides improvement over the ablation setting, selecting a mobility-related descriptor such as Doppler leads to significantly better prediction accuracy in highly dynamic OTFS channels.

The NMSE performance versus the prediction horizon is illustrated in Fig. $2$, where the horizon denotes the number of future OTFS frames to be predicted. As the prediction horizon increases, the accuracy of both methods degrades due to the reduced temporal correlation between the observed and future channel realizations. Nevertheless, the proposed physics-aware predictor consistently achieves lower NMSE across all horizons. For example, at a prediction horizon of 10 frames, the proposed method attains an NMSE of approximately $3\times10^{-3}$, whereas the ablation model yields about $3\times10^{-2}$. This result highlights the benefit of incorporating mobility-related physical information for improving LLM-based channel prediction in highly dynamic OTFS channels.
\begin{figure}[t]\label{fig:1}
\centering
        \includegraphics[width=3.4in]{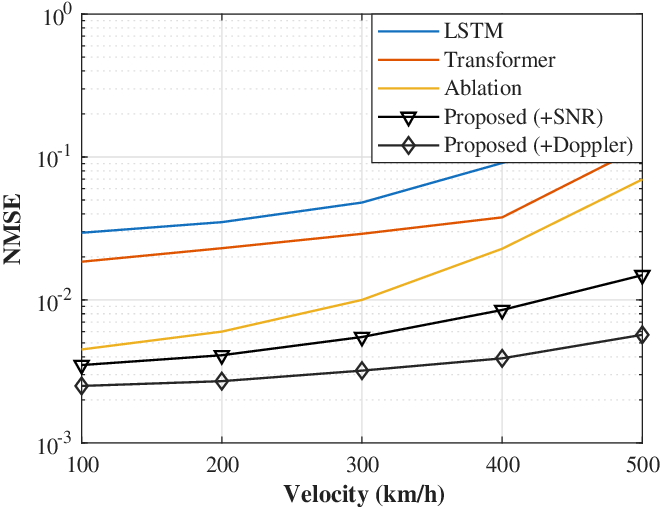}
     \caption{NMSE performance versus user velocity for different channel prediction methods.}
    \label{fig:1}
\end{figure}
\begin{figure}[t]\label{fig:2}
\centering
        \includegraphics[width=3.4in]{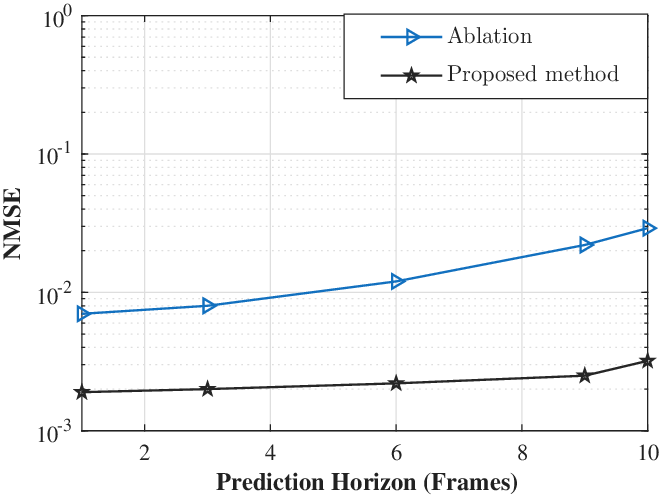}
     \caption{NMSE performance versus prediction horizon (number of future OTFS frames).}
    \label{fig:1}
\end{figure}

\section{Conclusion}\label{defin6}
In this paper, we presented an LLM-based framework for channel prediction in high-mobility OTFS communication systems, aiming to maintain accurate and timely channel knowledge in rapidly time-varying environments while reducing pilot overhead and improving spectral efficiency. First, a physics-aware prediction architecture was developed in which sequences of past OTFS channel coefficients are processed jointly with mobility-related physical parameters to capture both temporal channel correlations and mobility-induced variations. Second, the LLM backbone was utilized as a powerful sequence modeling tool to learn the complex temporal dependencies present in highly dynamic wireless channels. Simulation results demonstrate that the proposed framework effectively tracks the temporal evolution of OTFS channels under high mobility. Compared with conventional deep learning predictors and an ablation model that rely solely on historical channel observations, the proposed method using an appropriate physical descriptor (e.g., Doppler frequency), achieves consistently lower NMSE across different mobility conditions. These results highlight the advantage of incorporating physical channel knowledge into data-driven sequence modeling for reliable channel prediction. Future work will extend the proposed framework to more complex scenarios, including multi-user OTFS systems and heterogeneous propagation environments. 

\leavevmode%


\ifCLASSOPTIONcaptionsoff
  \newpage
\fi
\bibliographystyle{IEEEtran}
\bibliography{keylatex}
\end{document}